\documentclass[amsfonts, amssymb, amsmath, twocolumn, showkeys, nofootinbib, reprint,floatfix, prl, superscriptaddress]{revtex4-2}

\usepackage{amssymb}
\usepackage{amsmath}
\usepackage{mathtools}
\usepackage{bm}
\usepackage{color}
\usepackage{graphicx}
\usepackage[normalem]{ulem}
\usepackage[makeroom]{cancel}
\usepackage{units}
\usepackage{gensymb}
\usepackage{comment}
\usepackage{bbm}

\newcommand{\ve}{{\varepsilon}}
\newcommand{\bj}{{\bf j}}
\newcommand{\bk}{{\bf k}}
\newcommand{\br}{{\bf r}}
\newcommand{\bv}{{\bf v}}
\newcommand{\bE}{{\bf E}}
\newcommand{\bB}{{\bf B}}
\newcommand{\bOmega}{{\boldsymbol \Omega}}
\newcommand{\bmom}{{\bf m}}
\def \be{\begin{equation}}
\def \ee{\end{equation}}

\usepackage{hyperref}
\hypersetup{
  breaklinks = {true},
  citecolor = {blue},
  colorlinks = {true},
  linkcolor = {red},
  urlcolor = {blue},
}
\usepackage{lipsum}
\graphicspath{{../Figs/}}

\begin{document}

\title{Linear magneto-conductivity as a DC probe of time-reversal symmetry breaking}

\author{V. Sunko}
\thanks{V. S., C.L. and  M. V.  contributed equally to this work.}
\affiliation {Department of Physics, University of California, Berkeley, California 94720, USA}
\affiliation {Materials Sciences Division, Lawrence Berkeley National Laboratory, Berkeley, California 94720, USA}

\author{C. Liu}
\thanks{V. S., C.L. and  M. V.  contributed equally to this work.}
\affiliation {Department of Physics, University of California, Berkeley, California 94720, USA}

\author{M. Vila }
\thanks{V. S., C.L. and  M. V. contributed equally to this work.}
\affiliation {Department of Physics, University of California, Berkeley, California 94720, USA}
\affiliation {Materials Sciences Division, Lawrence Berkeley National Laboratory, Berkeley, California 94720, USA}

\author{I. Na}
\affiliation {Department of Physics, University of California, Berkeley, California 94720, USA}
\affiliation {Materials Sciences Division, Lawrence Berkeley National Laboratory, Berkeley, California 94720, USA}

\author{Y. Tang}
\affiliation {Department of Physics, University of California, Berkeley, California 94720, USA}

\author{V.~Kozii}
\affiliation {Department of Physics, Carnegie Mellon University, Pittsburgh, Pennsylvania 15213, USA}

\author{ S. M. Griffin }
\affiliation {Materials Sciences Division, Lawrence Berkeley National Laboratory, Berkeley, California 94720, USA}
\affiliation {Molecular Foundry, Lawrence Berkeley National Laboratory, Berkeley, California 94720, USA}

\author{J. E. Moore}
\email{jemoore@berkeley.edu}

\author{J. Orenstein}
\email{jworenstein@lbl.gov}

\affiliation {Department of Physics, University of California, Berkeley, California 94720, USA}
\affiliation {Materials Sciences Division, Lawrence Berkeley National Laboratory, Berkeley, California 94720, USA}

\date{\today}

\begin{abstract}
Several optical experiments have shown that in magnetic materials the principal axes of response tensors can rotate in a magnetic field. Here we offer a microscopic explanation of this effect, and propose a closely related  DC transport phenomenon --- an off-diagonal \emph{symmetric} conductivity linear in a magnetic field, which we refer to as linear magneto-conductivity (LMC). Although LMC has the same functional dependence on magnetic field as the Hall effect, its origin is fundamentally different: LMC requires time-reversal symmetry to be broken even before a magnetic field is applied, and is therefore a sensitive probe of magnetism. We demonstrate LMC in three different ways: via a tight-binding toy model, density functional theory calculations on MnPSe$_3$, and a semiclassical calculation. The third approach additionally identifies two distinct mechanisms yielding LMC: momentum-dependent band magnetization and Berry curvature. Finally, we propose an experimental geometry suitable for detecting LMC, and demonstrate its applicability using Landauer--B\"{u}ttiker simulations. Our results emphasize the importance of measuring the full conductivity tensor in magnetic materials, and introduce LMC as a new transport probe of symmetry.

\end{abstract}

\maketitle

In all materials, a current passing perpendicular to an applied magnetic field introduces a voltage transverse to both the current and the field \cite{Hall1879}. This phenomenon, called the Hall effect,  has been an invaluable tool for probing the concentration and character of charge carriers.
In magnetic materials an applied  field is not necessary to observe  the Hall effect: for example, the anomalous Hall effect (AHE) can be determined by the electronic structure and Berry curvature in momentum space \cite{Nagaosa2010}, whereas the topological Hall effect is induced in non-collinear magnetic structures by real-space Berry curvature \cite{nagaosa_topological_2013}.  Regardless of microscopic origins, any Hall effect requires time-reversal symmetry (TRS) to be broken~\cite{Chen2013,Smejkal2020}. 

The Hall effect arises from the anti-symmetric component of the conductivity tensor ($\sigma^a_{xy}=(\sigma_{xy}-\sigma_{yx})/2$), and is usually extracted from transverse voltage as a function of magnetic field ($H$), assuming implicitly that the  Hall effect is the only $H$-odd component of $\sigma$, i.e. that  the symmetric off-diagonal component $\sigma^s_{xy}=(\sigma_{xy}+\sigma_{yx})/2$ is $H$-even. In \emph{non-magnetic} materials the validity of this assumption is guaranteed by the Onsager's reciprocity relations, which encode microscopic time reversibility.

Here we show that this standard assumption is dangerous in magnetic systems: certain magnetic structures allow $\sigma^s_{xy}$ to have an $H$-odd component, which need not be small. A typical measurement cannot distinguish such response from the Hall effect, leading to possible misinterpretations of the measured signal. On the other hand, detecting a component of $\sigma^s_{xy}$ that is proportional to the applied field would provide evidence of TRS breaking that is complementary to the AHE, as the symmetry constraints for the two effects to be observed are distinct \cite{supp}. Measuring the full conductivity tensor $\sigma_{ij}$ as a function of magnetic field is therefore both a powerful probe of symmetry, and a necessity to correctly estimate the carrier concentration in magnetic materials. 

\begin{figure}[t]
\centering
\includegraphics[width= 0.86 \linewidth]{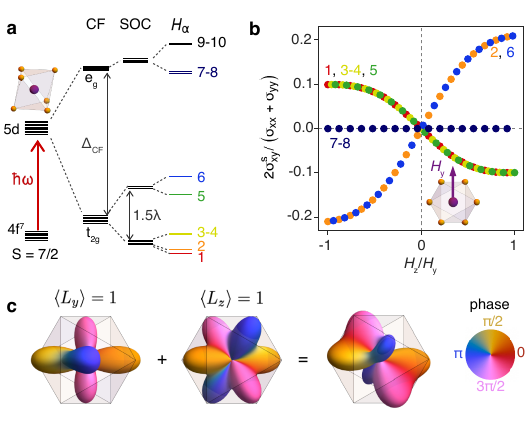}
\caption{ (a) The model considers a one electron $f\rightarrow d$ transition, with a degenerate initial state manifold and a final state manifold split by the three terms in Eq.~\ref{eq:Hamiltonian}: $\Delta_{CF}$ separates the $d$ manifold into states of $e_g$ and $t_{2g}$ symmetry. (b) The ratio of off-diagonal symmetric  conductivity to the average diagonal conductivity as a function of $H_z/H_y$, calculated for the different final states in Fig.~\ref{fig:Model}a.  (c) The orbital wave functions corresponding to $\left<L_{\alpha}^{\text{eff}}\right>=1$, for $\alpha=y, z, (z+y)/\sqrt{2}$.}
\label{fig:Model}
\end{figure}

The equivalent phenomenon at optical frequencies manifests as an $H$-linear rotation of principal optical axes, and  is referred to as off-diagonal linear magneto-birefringence (LMB). It can be used to determine the magnetic point group of a material, as has been reported several times \cite{Eremenko1986, Sunko2023}. Although the symmetry constraints for its existence are known \cite{Eremenko1987}, the microscopic understanding of the mechanism behind the principal axis rotation is still lacking. Furthermore, since the symmetry-allowed form of the conductivity tensor does not depend on the frequency, an analogous DC transport effect is in principle allowed. To the best of our knowledge, this DC phenomenon has not been explored theoretically nor experimentally. 

In this work, we take up this quest to examine the field-induced principal axis rotation in the DC limit, an effect we refer to as linear magneto-conductivity (LMC), and elucidate how LMC arises from the electronic structure of a magnetic material. We start from a microscopic understanding of the off-diagonal LMB in a system of localized orbitals in a magnetic field, and use the obtained insights to construct a tight-binding toy model which yields a nonzero DC LMC. Within this model the effect originates from a field-induced rotation of the Fermi surface (FS), resembling the rotation of principal optical axes that characterizes LMB. To gain insight into the physical origin of this effect, we derive the field-linear $\sigma^s_{xy}$ from semiclassical equations of motion and identify two mechanisms: Berry curvature and momentum-dependent band magnetization. Although TRS needs to be broken, neither of these mechanisms relies on the existence of a net magnetization, suggesting that LMC may be present in a wide range of magnetic materials. We propose an experimental geometry suitable for measuring the full conductivity tensor, and demonstrate it by a mesoscopic transport calculation. Finally, we use density-functional theory (DFT) calculations on an example material, MnPSe$_3$, to show that a large LMC is present beyond simple toy models, and needs to be seriously considered in experimental investigations.

We start by studying the origin of the optical effect. We develop a minimal model of EuCd$_2$P$_2$, a material in which off-diagonal LMB has been observed \cite{Sunko2023}, and use it to calculate the optical conductivity tensor by employing the Kubo formula \cite{supp}. For simplicity we consider an isolated Eu$^{2+}$ ion in an octahedral crystal field of $D_{3d}$ symmetry, and model a one-electron $f\rightarrow d$ transition (Fig.~\ref{fig:Model}a). The initial state configuration ($^8S_{7/2}$,  $L=0$, $S=7/2$) consists of seven degenerate spin-polarized $f$ states, which are not split by crystal field or spin-orbit coupling, since  $L=0$. The final state $5d$ manifold is split by a crystal field ($\Delta_{CF}\hat{H}_{CF}$), spin-orbit coupling ($\lambda$) and a Zeeman term induced by a magnetic field $\mathbf{H}$:
\begin{equation}\label{eq:Hamiltonian}
\hat{H}_{L}\left(\Delta_{CF},\lambda,\mathbf{H}\right)=\Delta_{CF}\hat{H}_{CF}+\lambda\mathbf{\hat{L}}\cdot{\mathbf{\hat{S}}}-\mu_B\mathbf{H}\cdot\big(2\mathbf{\hat{S}}+\mathbf{\hat{L}}\big),
\end{equation}
where $\mathbf{\hat{L}}$ and $\mathbf{\hat{S}}$ are orbital angular momentum (OAM) and spin operators, respectively, and $\mu_B$ is the Bohr magneton. The resulting energy spectrum, assuming $\Delta_{CF}\gg\lambda\gg H$, is shown in Fig.~\ref{fig:Model}a. The conductivity tensor \cite{supp}
can now be evaluated for transitions between the degenerate initial state manifold and each of the final states, calculated assuming an applied $\mathbf{H}$ within the $yz$ plane. This field captures the effect of an intrinsic magnetization $M_y$ that is tilted by an applied $H_z$.  We indeed find a non-zero $\sigma^s_{xy} \sim H_z/H_y$ (Fig.~\ref{fig:Model}b), but only for $t_{2g}$ final states, indicating the importance of OAM, which is quenched in the $e_g$ manifold. 

The response is caused by a subtle interference effect between wave functions which hold a non-zero OAM in the $y$ and $z$ directions ($L^{\text{eff}}_{y}$ and $L^{\text{eff}}_{z}$, Fig.~\ref{fig:Model}c). Both the amplitude and the phase of the wave functions have a non-trivial angular dependence; the former is caused by the CF which encodes spatial symmetries, and the latter by the field-induced OAM. Their interference results in the rotation of charge density, leading to the principal axis rotation. The effect is therefore expected for optical transitions in which either the initial or the final state supports a non-zero OAM, despite the lowering of symmetry by the CF, as is the case for $t_{2g}$ orbitals. 

\begin{figure}[t] 
\centering
\includegraphics[width=0.48\textwidth]{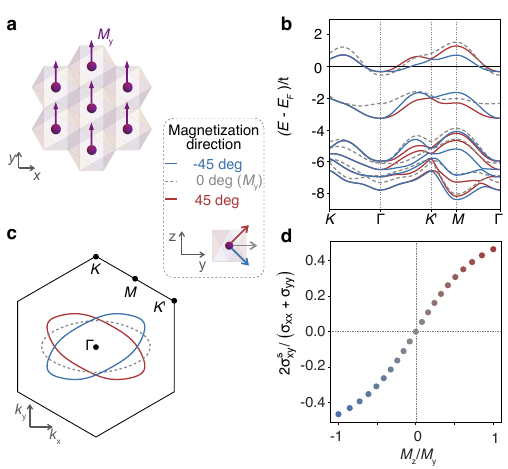}
\caption{Rotation of Fermi surfaces and change of conductivity in a tight binding model as a function of magnetization. (a) Top view of the triangular lattice of magnetic ions with moment ${\bf M}=(0,M_y,M_z)$ in the $y-z$ plane. The panels (b-d) study quantities when ${\bf M}$ points along $-45^\circ$, $0^\circ$, or $+45^\circ$ from the $y$ axis, represented by red, grey, and blue arrows. (b) The band dispersion along high symmetry path in the Brillouin zone (BZ) for the three orientations of ${\bf M}$. $t$ is the overall hopping energy scale and $E_F$ is the Fermi energy. (c) FS at the Fermi energy $E = E_F$ for the three orientations of ${\bf M}$. As ${\bf M}$ is oriented from $-45^\circ$ to $+45^\circ$, the FS rotates accordingly in a clockwise way. (d) The conductivity angle $2\sigma^s_{xy}/(\sigma_{xx}+\sigma_{yy})$ as a function of $M_z/M_y$. For (c) and (d) we assumed the filling at $E=E_F$.}
\label{fig:TightBinding}
\end{figure}

The above insight into the mechanism behind the off-diagonal LMB at optical frequencies suggests that an analogous effect in the DC limit might be achieved in a system with $t_{2g}$ orbitals at the Fermi level. We therefore construct such a tight-binding (TB) model, based on hoppings between local $d$ orbitals experiencing a strong octahedral crystal field splitting, and sitting on a  triangular lattice (Fig.~\ref{fig:TightBinding}a, $D_{3d}$ symmetry) \cite{supp}. The TB Hamiltonian has the form
\begin{equation}\label{eq:TBHam}
\hat{H}_{TB}= \hat{H}_{L} +\sum_{{\bf rr'},\ell\ell',\sigma\sigma'}t_{\mathbf{r,r'}\!,\ell\ell'\!,\sigma\sigma'} d^\dag_{\mathbf{r},\ell,\sigma} d_{{\bf r'},\ell',\sigma'} ,
\end{equation}
where $\ell$ ($\sigma$) label the $d$ orbitals (spins). The first term, $\hat{H}_{L} = \sum_{\mathbf{r}} d^\dag_{\mathbf{r}} \hat{H}_L(\Delta_{\text{CF}},\lambda,{\bf M})d_{\mathbf{r}}$, is equivalent to the local term in Eq.~\eqref{eq:Hamiltonian}, with magnetization ${\bf M}$ in place of an external field, while the second term describes nearest neighbor hopping between such orbitals; the hopping matrix elements $t_{\bm{r}\bm{r}'\!,\ell\ell'\!,\sigma\sigma'}$ are obtained from the Slater-Koster parameterization \cite{supp,PhysRev.94.1498}.

We plot in Fig.~\ref{fig:TightBinding}b the band structure of the $t_{2g}$-derived bands calculated for three orientations of magnetization ${\bf M}$: along $y$ axis and $\pm 45^\circ$ away from the $y$ axis in the $yz$ plane. The bands are clearly modified by the magnetization. More insights can be gleaned from the effect of ${\bf M}$ on the FS: the FS is elliptical for all three orientations of ${\bf M}$, and it rotates in the $k_xk_y$ plane as the magnetization ${\bf M}$ rotates in the $yz$ plane.

The rotation of the FSs induced by the magnetization has profound consequences on transport.  The conductivity tensor depends on the geometry of an elliptical FS as:
\begin{equation}
\sigma = \sigma_0 \mathbbm{1} + \sigma_0 \mathcal{E}^2\!\begin{pmatrix} -\cos 2\mathsf{\theta} & \sin 2\mathsf{\theta}\\\sin 2\mathsf{\theta} &  \cos 2 \mathsf{\theta}\end{pmatrix},
\end{equation}
where $\mathbbm{1}$ is the unit matrix, and $\mathsf{\theta}$ and $\mathcal{E}$ denote the orientation of the major axis of the ellipse and its eccentricity, respectively. The isotropic part $\sigma_0\mathbbm{1}$ denotes the conductivity in absence of ${\bf M}$, as enforced by the threefold rotation and mirror symmetries \cite{supp}. Importantly, the principal axes of the conductivity tensor correspond to the major and minor axes of the ellipse, and a nonzero FS rotation $\theta$ induces a nonzero off-diagonal conductivity. Since $\sigma_{xy}$ is directly proportional to $\sin 2\mathsf{\theta}$, we expect the opposite sign of  $\sigma_{xy}$ for the two FSs shown in Fig.~\ref{fig:TightBinding}c, calculated for the magnetization along $\pm\unit[45]{\degree}$.

We confirm the above picture by calculating $\sigma^s_{xy}$ as a function of the orientation of ${\bf M}$ using the Drude formula. In Fig.~\ref{fig:TightBinding}d we plot the ratio $2\sigma^s_{xy}/(\sigma_{xx}+\sigma_{yy})$, which is independent of the relaxation time $\tau$, as a function of  $M_z/M_y$. Importantly,  $\sigma^s_{xy}$ is an odd function of $M_z$, and exhibits linear dependence at small $M_z$, just like the (anti-symmetric) Hall conductivity. Therefore, extracting the $H$-linear component of the transverse voltage measured in a single geometry, as is usually done, is not sufficient to isolate the Hall signal in a magnetic material, and independent measurements of $\sigma_{xy}$ and $\sigma_{yx}$ are needed.

Our TB model shows that a magnetization can induce LMC, but it does not prove that magnetization is \emph{necessary} for this effect to take place; this is reminiscent of the AHE, which can be induced by a magnetization, but it does not require it. To explore this analogy, we have derived the expression for conductivity $\sigma^s_{ij}$ of a two-dimensional (2D) system up to the linear order in ${\bf B} = B_z\hat{z}$ within the semiclassical approximation \cite{RevModPhys.82.1959}. Note that in semiclassics the electrons are affected by the field  ${\bf B} = \mu_0 ({\bf H} + {\bf M})$. 
The full derivation is  given in \cite{supp}, including corrections due to a field-induced shift of the chemical potential, while the final expression relevant for the discussion here is \cite{footnote1}: 
\begin{equation}
\begin{aligned}\label{eq:Sigma}
&\sigma^s_{xy} =  B_z \frac{\tau e^2}{\hbar} \int \frac{d^2k}{(2\pi)^2}    \\
&\,\, \times \left[\frac{1}{2}(v^0_x\partial_y m_{\mathbf{k}}+v^0_y\partial_xm_{\mathbf{k}})
-m_{\mathbf{k}} v^0_{xy}+ ev^0_xv^0_y\Omega_\bk \right]f_0',
\end{aligned}
\end{equation}
where $\tau$ is the relaxation time and $e$ the elementary charge. Inside the integral, $f'_0$ is the derivative of the Fermi function $f_0(\varepsilon_{0,\mathbf{k}})$ with respect to the single-particle energy $\varepsilon_{0,\mathbf{k}}$, $v_i^0, i=x,y$ are the components of the Fermi velocity and we defined their derivative with respect to momentum as $v_{ij}^0 = \partial_i v^0_j=\partial_j v^0_i$, $\Omega_{\mathbf{k}}$ is the Berry curvature, and ${\bf m}_{\mathbf{k}} = m_{\mathbf{k}}\hat{z}$  is the magnetization of the conduction electrons, defined by the shift of the single-particle energy in an external field, $\widetilde{\varepsilon}_{\bf k}=\varepsilon_{0,{\bf k}}-{\bf B}\cdot {\bf m}_{\mathbf{k}}$~\cite{PhysRevLett.95.137205}.
We find two distinct types of contributions at linear order in $B_z$: due to magnetization and Berry curvature.  The first three terms in the integral describe a ``magnetization" contribution; it is further divided into a ``velocity'' part (first two terms) and a ``band curvature'' part (third term). The fourth term, on the other hand, describes a ``Berry curvature'' contribution. 
The physical origin and behavior of these terms differ.

The ``magnetization'' contribution arises due to 
momentum-dependent changes of magnetization and band velocities. Note that if the magnetization is constant or linear in momentum across the momentum space, 
the ``velocity'' part of the ``magnetization'' contribution  to LMC vanishes, suggesting that it is expected to be significant in systems with strongly mixed orbital character at the Fermi level. This finding is consistent with the intuition obtained from the local-orbital picture (Fig.~\ref{fig:Model}): orbital mixing, promoted by crystal field and spin--orbit coupling, leads to non-zero expected value of OAM, as well as the momentum-depended response to a magnetic field. On the other hand, the ``band curvature'' part of the ``magnetization'' contribution to LMC vanishes if the band disperses linearly in momentum space. 
 The ``Berry curvature'' contribution is proportional to the Berry curvature $\Omega_\bk$ multiplied by the components of the Fermi velocity, and it arises from Berry phase correction to the phase space volume \cite{Xiao2005}. This term can in principle give a non-zero contribution even if $\Omega_\bk$ is constant across the momentum space. Symmetry such as a twofold rotation or a mirror reflection can promote the LMC effect. 

We note that although AHE can also arise from $\Omega_\bk$, its functional form differs substantially from that of LMC:
\begin{equation}\label{eq:AHE}
\sigma^{\text{AHE}}_{xy} =  -\frac{e^2}{\hbar}\int \frac{d^2k}{(2\pi)^2}\, \Omega_\bk\,f_0.
\end{equation}
First of all, $\sigma^{\text{AHE}}_{ij}$ does not depend on the Fermi velocity or the relaxation time. Second, the integral in Eq.~\eqref{eq:AHE} is taken over all occupied states (characterized by non-zero Fermi function $f_0$), while the one in Eq.~\eqref{eq:Sigma} is taken over the vicinity of FS (characterized by non-zero $f'_0$). These differences emphasize that the AHE and LMC are two distinct probes of TRS breaking, as already deduced from symmetry considerations \cite{supp}.

\begin{figure}[t] 
\centering
\includegraphics[width=0.48\textwidth]{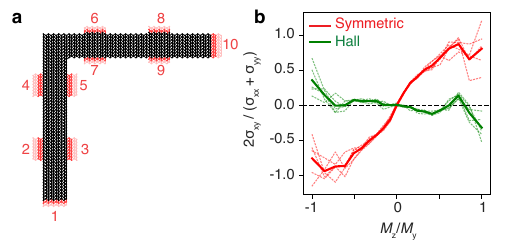}
\caption{(a) L-shape device geometry. The scattering region (leads) is denoted by the black (red) lattice. Leads 1 and 10 are the source and drain electrodes, while leads 2-8 are voltage probes. (b) Conductivity angle for the symmetric (red) and Hall (green) conductivity as a function of $M_z/M_y$. The calculations are done based on the tight binding model (Eq.~\eqref{eq:TBHam}, Fig.~\ref{fig:TightBinding}), with a Fermi level of $-1.95$. Solid line is the average value performed over 5 disorder realizations (shown in dashed, paler color).}
\label{fig:LB}
\end{figure}

Having confirmed that DC linear magneto-conductivity is not only allowed by symmetry, but also expected based on microscopic arguments, we now turn towards two experimental aspects: we propose an experimental geometry suited for the detection of LMC, and demonstrate a sizable expected LMC in a realistic material by first-principles calculations. 

Long Hall bars are typically used for transport measurement to allow the current to become homogeneous between the measurement contacts, and therefore reject effects from the injection region. Simply exchanging current and voltage contacts in a single device, which would theoretically probe $\sigma_{xy}$ and $\sigma_{yx}$, would introduce different artifacts in the two measurements. Another strategy might be to fabricate two devices, with the bars along two orthogonal crystal directions. However, this would introduce sample-to-sample variation, which may be a serious limitation since LMC and the Hall effect exhibit different functional dependence on the scattering rate.  This leads us to propose an L-shaped device,  illustrated in Fig.~\ref{fig:LB}(a), to independently measure $\sigma_{xy}$ and $\sigma_{yx}$ in the same geometry and in the same crystal, allowing for robust extraction of $\sigma^a$ and $\sigma^s$. Such devices can be structured from single crystals using focused ion beam milling, or created in thin flakes using lithographic techniques.  

To demonstrate this approach, we perform Landauer--B\"{u}ttiker calculations in the $L$-shaped geometry \cite{supp}, assuming bands given by the TB model above (Eq.~\eqref{eq:TBHam}, Fig.~\ref{fig:TightBinding}). The resulting $2\sigma^s_{xy}/\left(\sigma_{xx}+\sigma_{yy}\right)$ is plotted as a function of the orientation of ${\mathbf M}$ in Fig.~\ref{fig:LB}b. The qualitative agreement of the Landauer--B\"{u}ttiker calculations with the Drude formula (see Fig.~\ref{fig:TightBinding}d) is excellent. For comparison, we also plot $2\sigma^a_{xy}/\left(\sigma_{xx}+\sigma_{yy}\right)$, from which we see that $\sigma^s_{xy}$ is larger than the Hall conductivity for our choice of parameters, demonstrating that this regime can indeed be reached. We note that the ratio  $\sigma^s_{xy}/\sigma^a_{xy}$ depends on the scattering time, and becomes larger with increasing purity. 

\begin{figure}[t] 
\centering
\includegraphics[width=0.45\textwidth]{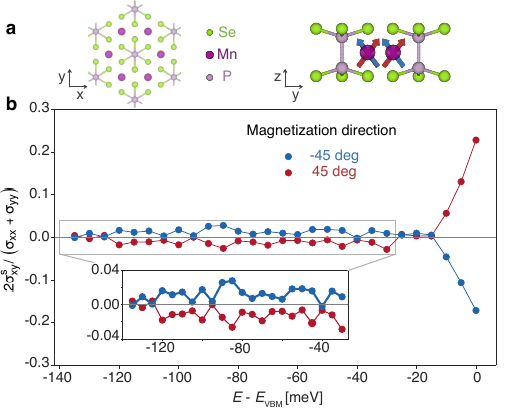}
\caption{(a) Top view and side view of the MnPSe$_3$ crystal structure. The red and blue arrows indicate the two magnetization directions used in the calculations. (b) Conductivity angle as a function of energy for magnetization orientation along $\pm45\degree$ with respect to the $z$ axis in the $yz$ plane.}
\label{fig:DFT}
\end{figure}

Finally, we expand our TB model and semiclassical analysis by evaluating LMC from DFT calculations of a real material, a monolayer of MnPSe$_3$ \cite{supp}. This is a member of the family of magnetic 2D transition metal phosphorous tricalcogenides that have gathered increasing attention recently for their wealth of interesting magnetic, optical and topological properties \cite{Chittari2016, samal_two-dimensional_2021, Natalya2023, Na2023}. MnPSe$_3$ adopts the trigonal $R\overline{3}$ space group where a layer of magnetic Mn atoms sits on a honeycomb lattice surrounded by PSe$_3$ layers above and below it (Fig.~\ref{fig:DFT}a). Its ground state magnetic order is antiferromagnetic -- here we consider the metastable ferromagnetically ordered case, as has been previously investigated~\cite{Na2023, Natalya2023}. We choose to calculate LMC in MnPSe$_3$ because it has the same point group symmetry as our toy model ($D_{3d}$, Fig.~\ref{fig:TightBinding}), and its low-energy electronic structure contains $t_{2g}$ orbitals of Mn.

We performed DFT calculations for two orientations of Mn magnetic moments within the $yz$ plane (Fig.~\ref{fig:DFT}a, right panel), and used the Drude formula to evaluate $2\sigma^s_{xy}/\left(\sigma_{xx}+\sigma_{yy}\right)$ as a function of the Fermi level. As can be seen in Fig. \ref{fig:DFT}b, a non-zero $\sigma^s_{xy}$ is predicted across a wide range of energies. Its sign depends on the orientation of the magnetic moment, thus validating our understanding of the LMC and its presence in MnPSe$_3$. An important finding is the magnitude of the effect. For a Fermi energy close to the valence band maximum (VBM) $\sigma_{xy}^s$ reaches values as large as $25\%$ of the diagonal conductivity. This dramatic enhancement occurs because the small FSs close to the VBM can be significantly perturbed by the magnetization, and emphasizes the importance of measuring the full conductivity tensor in studies of magnetic semiconductors and (semi)metals. Even without this enhancement, $\sigma_{xy}^s$ reaches $1\%-3\%$ of the diagonal conductivity over a wide range of Fermi energies, which is substantial compared to typically observed Hall effect values. Indeed, both anomalous Hall effect \cite{yang_giant_2020} and topological Hall effect \cite{kurumaji_skyrmion_2019}  are considered ``giant'' when they correspond to a few percent of longitudinal conductivity. 

To conclude, the impact of our work is threefold. First of all, we have identified a microscopic mechanism responsible for the rotation of principal optical axes in a magnetic field. Furthermore, we demonstrated a fundamentally new transport effect, referred to as linear magneto-conductivity, using three complementary approaches: a tight-binding toy model, DFT calculations and a semiclassical calculation. The latter is particularly powerful, as it identifies two distinct mechanisms yielding LMC: momentum dependent band magnetization, and Berry curvature. Finally, we have shown the importance of measuring the full conductivity tensor in magnetic materials, both because the Hall effect can otherwise be misidentified, and because such measurements serve as a powerful probe of symmetry. We hope that our results inspire new ways of probing the effect of time-reversal symmetry breaking on conduction electrons in future measurements.

\begin{acknowledgments}
M.V. is grateful to Stephen R. Power for helpful discussions.
V.S. is supported by the Miller Institute for Basic Research in Science, UC Berkeley.
C.L. acknowledges fellowship support from the Gordon and Betty Moore Foundation through the Emergent Phenomena in Quantum Systems (EPiQS) program.
M.V. was supported as part of the Center for Novel Pathways to Quantum Coherence in Materials, an Energy Frontier Research Center funded by the US Department of Energy, Office of Science, Basic Energy Sciences. J.M. acknowledges support from the Quantum Materials program under the Director, Office of Science, Office of Basic Energy Sciences, Materials Sciences and Engineering Division, of the U.S. Department of Energy, Contract No. DE-AC02-05CH11231. J.O received support from the Gordon and Betty Moore Foundation's EPiQS Initiative through Grant GBMF4537 to J.O. at UC Berkeley. I. N. and S. G. were supported by the  U.S. Department of Energy, Office of Science, Office of Basic Energy Sciences, Materials Sciences and Engineering Division under Contract No.DE-AC02-05-CH11231 within the Theory of Materials program.Computational resources were provided by the National Energy Research Scientific Computing Center and the Molecular Foundry, DOE Office of Science User Facilities supported by the Office of Science, U.S. Department of Energy under Contract No. DEAC02-05CH11231.The work performed at the Molecular Foundry was supported by the Office of Science, Office of Basic Energy Sciences of the U.S. Department of Energy, under the same contract No.

 \end{acknowledgments}

\bibliography{bib}

\begin{thebibliography}{41}
\bibitem [{foo()}]{footnote1}%
  \BibitemOpen
  \bibinfo {note} {{Note that if assuming the electron density is fixed upon
  turning on magnetic field, an additional term may exist as the result of
  renormalization of chemical potential. Such a term vanishes in systems with
  mirror symmetry (which are the ones considered in this work). For more
  detail, see discussions in \cite{supp}.}}\BibitemShut {Stop}%
\end{thebibliography}

\appendix

\newpage

\noindent

\onecolumngrid

\counterwithout{equation}{section}

\renewcommand\theequation{S\arabic{equation}}
\renewcommand\thefigure{S\arabic{figure}}
\renewcommand\bibnumfmt[1]{[S#1]}
\setcounter{equation}{0}
\setcounter{figure}{0}
\setcounter{enumiv}{0}

\section*{Supplementary Material for ``Linear magneto-conductivity as a DC probe of time-reversal symmetry breakin''}

\section{Symmetry conditions required for linear magneto-conductivity and AHE}\label{sec:symmetries}

We discuss in the main text that the anomalous Hall effect and linear magneto-conductivity are distinct probes of symmetry. To emphasize this point further,  in table~\ref{tab:symm} we list the symmetries that allow ($\checkmark$) or prohibit ($\times$) the two effects. In particular, AHE is prohibited by mirror symmetries $m_x$ and $m_y$, which allow for LMC. In contrast, LMC requires the $z$-axis to have at most the $C_{2z}$ rotational symmetry, while AHE is not limited by rotational symmetry. 

\begin{table}[!thb]
\begin{tabular}{ccc}
\hline\hline
Symmetry            & AHE          & LMC          \\ \hline
$\mathcal{T}$       & $\times$     & $\times$     \\
$\mathcal{P}$       & $\checkmark$ & $\checkmark$ \\
$\mathcal{PT}$      & $\times$     & $\times$     \\
$m_x$               & $\times$     & $\checkmark$ \\
$m_y$               & $\times$     & $\checkmark$ \\
$m_z$               & $\checkmark$ & $\checkmark$ \\
$m'_x$              & $\checkmark$ & $\times$     \\
$m'_y$              & $\checkmark$ & $\times$     \\
$m'_z$              & $\times$     & $\times$     \\
$\mathcal{C}_{2z}$  & $\checkmark$ & $\checkmark$ \\
$\mathcal{C}_{3z}$  & $\checkmark$ & $\times$     \\
$\mathcal{C}_{4z}$  & $\checkmark$ & $\times$     \\
$\mathcal{C}'_{2z}$ & $\checkmark$ & $\times$  \\
$\mathcal{C}'_{3z}$ & $\checkmark$ & $\times$     \\
$\mathcal{C}'_{4z}$ & $\checkmark$ & $\checkmark$\\
\hline\hline
\end{tabular}\caption{Symmetry requirements for the anomalous Hall effect (AHE) and linear magneto-conductivity (LMC). $\checkmark$ and $\times$ denote that the effect is and is not allowed by the corresponding symmetry, respectively.} \label{tab:symm}
\end{table}

\section{Calculation of optical conductivity}\label{sec:optics}

Optical conductivity is calculated using the standard Kubo formula: 

\begin{equation}\label{eq:Kubo_position}
\sigma_{\alpha\beta}\left(\omega\right)=-\frac{\imath e^{2}}{V\hbar}\times\sum_{n,m}\left(f_{m}-f_{n}\right)\left(E_{m}-E_{n}\right)\frac{\left\langle n\left|\hat{\alpha}\right|m\right\rangle \left\langle m\left|\hat{\beta}\right|n\right\rangle }{\hbar\left(\omega+\imath\eta\right)-\left(E_{m}-E_{n}\right)},
\end{equation}
where $\hat{\alpha}$ and $\hat{\beta}$ are position operators ($\alpha, \beta=x,y,z$), $\omega$ is energy, $n$ and $m$ states, $f_n$ and $f_m$ their respective occupations, $E_n$ and $E_m$ their energies, and $\eta$ is a small damping. Initial states are taken to be seven degenerate spin-polarized  $f$-orbitals, while the final states are $d$-orbitals split by a trigonal crystal field and spin-orbit coupling.

\section{Semiclassical calculation}

In this section, we derive the most generic semiclassical expression for linear magnetoconductivity, assuming that the Fermi energy lies within a single non-degenerate band. We adopt the approach discussed in, e.g., Refs.~\cite{Morimoto2016,Kozii23}. 

We start with the semiclassical equations of motion having the form~\cite{Xiao2010}
\begin{align}
\hbar \dot \br  &= \nabla_\bk \widetilde\ve_\bk - \hbar \dot \bk \times {\bOmega}_{\bk}, \nonumber \\ 
\hbar \dot \bk &= -e \bE - e \dot\br \times \bB,
\end{align}
where ${\boldsymbol \Omega}_\bk = i \langle \nabla_{\bk} u_{\bk} | \times \nabla_{\bk} u_{\bk}    \rangle$ is the Berry curvature. The quasiparticle dispersion relation is modified in the presence of an external magnetic field $\bB$: 
\be
\widetilde\ve_\bk = \ve^0_\bk - {\bf m}_\bk \cdot \bB,
\ee 
where $\ve^0_\bk$ is the bare band energy at $\bB=0$, $H_\bk |u_\bk\rangle = \ve^0_\bk |u_\bk\rangle$, and the orbital magnetic moment equals
\be 
{\bf m}_\bk = -i \frac{e}{2\hbar} \langle \nabla_{\bk} u_{\bk} \left| \times \left(H_\bk - \ve_\bk^0\right)\right| \nabla_{\bk} u_{\bk}    \rangle .
\ee

The solution of the equations of motion reads as
\begin{align}
\dot \br &= \frac1{\hbar D_\bk} \left\{ \nabla_\bk \widetilde\ve_\bk + e \bE \times \bOmega_\bk  + \frac{e}{\hbar} \bB \left( {\bOmega}_\bk \cdot \nabla_\bk \widetilde\ve_\bk  \right) \right\}, \nonumber \\ \dot \bk &= \frac1{\hbar D_\bk}\left\{  -e \bE - \frac{e}{\hbar} \nabla_\bk \widetilde\ve_\bk \times \bB - \frac{e^2}{\hbar} (\bE \cdot \bB) {\bOmega}_\bk  \right\}, \label{Eq:EOMresolved}
\end{align} 
where the factor $D_\bk = 1 + (e/\hbar) (\bB \cdot {\bOmega}_\bk)$ accounts for the density of states modification~\cite{Xiao2005}. 

The current density in a uniform system equals 
\be  
{\bf j} = -e \int \frac{d^dk}{(2\pi)^d} D_\bk \dot \br f = - \frac{e}{\hbar} \int \frac{d^dk}{(2\pi)^d} \left[ \nabla_\bk \widetilde\ve_\bk + e \bE\times{\bOmega}_\bk + \frac{e}{\hbar} \bB (\nabla_\bk \widetilde\ve_\bk \cdot {\bOmega}_\bk) \right] f, \label{SMEq:jtotsc}
\ee 
where $d$ is the dimensionality of the system. The semiclassical distribution function $f$ satisfies the conventional kinetic equation. We assume the relaxation time approximation for simplicity, resulting in 
\be 
\frac{\partial f}{\partial t} + \dot \bk \cdot\nabla_\bk f = - \frac{f-f_0}{\tau},
\ee 
where $\dot \bk$ is given by Eq.~\eqref{Eq:EOMresolved}, $\tau$ is the relaxation time, and $f_0=f_0(\widetilde\ve_\bk)$ is the equilibrium Fermi-Dirac distribution, which at zero temperature merely equals $f_0=\Theta(\mu - \widetilde\ve_\bk)$. We emphasize that $f_0$ depends on the modified quasiparticle energy $\widetilde\ve_\bk$ rather than on the bare band energy $\ve_\bk^0$.  Assuming monochromatic electric field, 
\be
\bE(t) = \bE_\omega e^{-i\omega t} + \bE_{\omega}^* e^{i\omega t},
\ee 
the distribution function equals
\be  
f(t) = f_0 + f_1 e^{-i \omega t} +  f_1^* e^{i \omega t} + \ldots,
\ee 
where $f_1 \propto E = |\bE_\omega|$. Physical electric field $\bE(t)$ is real, implying that $\bE_{-\omega} = \bE_{\omega}^*$. The kinetic equation for $f_1$ then becomes
\be
D_\bk \left(i\omega - \frac1\tau\right)  f_1 = -  \frac{e}{\hbar}\left( \bE_\omega + \frac{e}{\hbar} (\bE_\omega \cdot \bB){\bOmega}_\bk \right)\cdot \nabla_\bk f_0 - \frac{e}{\hbar^2} \nabla_\bk f_1 \cdot \nabla_\bk \widetilde\ve_\bk \times \bB.
\ee
At small magnetic fields, the solution of this equation up to linear order in $\bB$ equals
\be
f_1(\bk) \approx \frac{e \tau}{\hbar D_\bk (1 - i\omega\tau)} \left\{   \left( \bE_\omega + \frac{e}{\hbar} (\bE_\omega \cdot \bB){\bOmega}_\bk \right)\cdot \nabla_\bk f_0(\widetilde\ve_\bk) + \frac{e \tau}{\hbar^2(1- i \omega\tau)}  \left(\nabla_\bk \widetilde\ve_\bk \times \bB\right) \cdot \nabla_\bk \left( \bE_\omega \cdot \nabla_\bk f_0(\widetilde\ve_\bk) \right)   \right\},
\ee
and we explicitly restored the argument of $f_0$. Expanding further terms with $\widetilde\ve_\bk = \ve^0_\bk - {\bf m}_\bk \cdot \bB$ up to linear order in ${\bf m}_\bk \cdot \bB$ and introducing bare band velocity $\bv^0_\bk = \nabla_\bk \ve^0_\bk/\hbar$, we find
\begin{align} \label{SMEq:f1}
&f_1(\bk) \approx \frac{e\tau}{\hbar(1 - i\omega \tau)}   \left\{  \hbar \left(\bE_\omega\cdot \bv^0_\bk\right)f_0'\left(\ve^0_\bk\right) - e(\bB \cdot \bOmega_\bk)(\bE_\omega \cdot \bv^0_\bk)f_0'\left(\ve^0_\bk\right) - \left(\bE_\omega \cdot \nabla_\bk (\bB \cdot \bmom_\bk)  \right)f_0'\left(\ve^0_\bk\right)  \right.  \\  &+ e(\bE_\omega \cdot \bB)(\bOmega_\bk \cdot \bv^0_\bk)f_0'\left(\ve^0_\bk\right) - \left. \hbar(\bE_\omega \cdot \bv^0_\bk)(\bB \cdot \bmom_\bk)f_0''\left(\ve^0_\bk\right)   + \frac{e \tau}{1-i\omega\tau} f_0'\left(\ve^0_\bk\right) \left( \bv^0_\bk \times \bB \right)  \cdot \nabla_\bk \left( \bE_\omega \cdot \bv^0_\bk   \right) \right\}. \nonumber
\end{align}

Importantly, $f_0'$ and $f_0''$ now have argument $\ve^0_{\bk}$, and we defined $f_0'(\ve) = \partial f_0\left(\ve\right)/\partial \ve$ and $f_0''(\ve) = \partial^2 f_0\left(\ve\right)/\partial \ve^2$.

From Eq.~\eqref{SMEq:jtotsc}, we find that the linear in $\bE$ part of the current equals $\bj_1 e^{-i\omega t} + \bj_1^* e^{i\omega t}$, with 
\be  
\bj_1 = -\frac{e}{\hbar} \int \frac{d^dk}{(2\pi)^d}\left\{\left[  \nabla_\bk \widetilde\ve_\bk + \frac{e}{\hbar}\left(\nabla_\bk \widetilde\ve_\bk \cdot \bOmega_\bk\right)\bB   \right]f_1(\bk) + e \bE_\omega\times \bOmega_\bk f_0\left(\widetilde\ve_\bk\right)  \right\}.
\ee 
Expanding again up to the linear order in $\bB$, we find
\be  
\bj_1 = \bj_1^{(0)} + \bj_1^{(1)},  
\ee 
where $\bB-$independent part is given by
\be  
\bj_1^{(0)} = - \frac{e^2}{\hbar} \int\frac{d^dk}{(2\pi)^d}\left\{ f_0\left(\ve_\bk^0\right) \bE_\omega \times \bOmega_\bk  + \frac{\hbar\tau f_0'\left(\ve^0_\bk\right)}{1 - i\omega \tau} \left(\bE_\omega \cdot \bv^0_\bk\right)  \bv^0_\bk   \right\}, \label{SMEq:j10}
\ee 
and linear in $\bB$ contribution has form
\begin{align}
&\bj_1^{(1)} = \frac{e^2}{\hbar} \int\frac{d^dk}{(2\pi)^d}(\bB\cdot \bmom_\bk)\left( \bE_\omega \times \bOmega_\bk \right) f_0'\left(\ve_\bk^0\right) + \\ &+  \frac{e^2 \tau}{\hbar (1-i\omega\tau)}\int\frac{d^dk}{(2\pi)^d} f_0'\left(\ve^0_\bk\right)\left\{ \left(\bE_\omega\cdot \bv^0_\bk\right)\nabla_\bk(\bB\cdot \bmom_\bk) + \bv^0_\bk (\bE_\omega \cdot \nabla_\bk (\bB \cdot \bmom_\bk)) +  \right. \nonumber \\ &+ \left. e \left[ \bv^0_\bk (\bB\cdot \bOmega_\bk)\left(\bE_\omega\cdot \bv^0_\bk\right) - \bv^0_\bk (\bE_\omega\cdot \bB) (\bOmega_\bk\cdot\bv^0_\bk) - \bB \left(\bE_\omega \cdot \bv^0_\bk \right) \left(\bOmega_\bk \cdot \bv^0_\bk\right) \right] \right\}  + \nonumber  \\ &+ \frac{e^2 \tau}{1-i\omega\tau}\int\frac{d^dk}{(2\pi)^d}  \bv^0_\bk\left\{ f_0''\left(\ve_\bk^0\right) \left(\bE_\omega\cdot \bv^0_\bk\right)(\bB\cdot \bmom_\bk) - \frac{e\tau}{\hbar(1-i\omega\tau)}f_0'\left(\ve_\bk^0\right)\left(\bv^0_\bk\times \bB\right)\cdot \nabla_\bk\left(\bE_\omega\cdot \bv^0_\bk\right) \right\}.\nonumber
\end{align}

The first term in this expression corresponds to the anomalous Hall effect correction due to finite magnetic field. This term is fully antisymmetric and thus does not contribute to $\sigma_{ij}^s$. The last term is entirely due to Lorentz force and is not related to the geometry of the band structure. It may have both symmetric and anti-symmetric components that we specify below. Finally, in $d=2$ which is the main focus of this paper, both $\bOmega_\bk$ and $\bmom_\bk$ are pointing in the $z-$direction, hence, the two terms containing  $\left(\bOmega_\bk \cdot \bv^0_\bk\right)$   are equal to 0.

So far we only considered the grand canonical ensemble, i.e., assumed that the chemical potential is fixed and does not depend on $\bB$. If we instead focus on the canonical ensemble and assume the fixed density of particles $n$,  we find that the chemical potential is renormalized as $\mu \to \mu - \delta \mu$. The correction $\delta \mu$ due to finite magnetic field is found from the condition 
\be  
n = \int \frac{d^d k}{(2\pi)^d}D_\bk f_0\left(\ve_\bk^0 + \delta \mu - \bB \cdot \bmom_\bk\right) = \int \frac{d^d k}{(2\pi)^d} f_0\left(\ve_\bk^0 \right)
\ee 
and equals to the leading order in $\bB$
\be  
\delta \mu = \delta \mu_\bmom + \delta \mu_\bOmega = \frac1{N_0} \int \frac{d^d k}{(2\pi)^d} \left\{ -(\bB \cdot \bmom_\bk) f_0'\left(\ve_\bk^0\right) + \frac{e}{\hbar} (\bB \cdot \bOmega_\bk) f_0\left(\ve_\bk^0\right) \right\}, \qquad N_0 = - \int \frac{d^d k}{(2\pi)^d} f_0'\left(\ve_\bk^0\right).
\ee 
At zero temperature,
\be  
N_0 = \int \frac{d^d k}{(2\pi)^d} \delta \left(\ve_\bk^0 - \mu  \right),
\ee
which is the zero-field density of states per unit volume at the Fermi level.

The corresponding correction to the current comes from the zero-order contribution $\bj_1^{(0)}$, Eq.~\eqref{SMEq:j10}, and equals

\be  
\delta \bj_1^{(\mu)} = -\frac{e^2}{\hbar} \delta \mu \int \frac{d^d k}{(2\pi)^d} \left\{ f'_0\left(\ve_\bk^0\right) \bE_\omega \times \bOmega_\bk  + \frac{\hbar\tau f_0''\left(\ve^0_\bk\right)}{1 - i\omega \tau} \left(\bE_\omega \cdot \bv^0_\bk\right)  \bv^0_\bk   \right\}.
\ee 
Again, the first term is the correction to the anomalous Hall effect, which is antisymmetric. The second term is the correction to the Drude conductivity: it is symmetric and contributes to $\sigma_{ij}^s$. The information about the band structure geometry is hidden in the correction $\delta \mu$.

Focusing on the case $d=2$, we have $\bOmega_\bk = \Omega_\bk \hat z$ and $\bmom_\bk = m_\bk \hat z$. Assuming further that $\bB = B_z \hat z$ and extracting the symmetric part of the conductivity tensor, $\sigma_{ij}^s = \left(\sigma_{ij} + \sigma_{ji}\right)/2$, we find 
\begin{align}\label{2Dsigma}
\sigma_{ij}^s &= \sigma_{ij}^D + \frac{e^2 \tau B_z}{\hbar(1-i\omega\tau)} \int \frac{d^2k}{(2\pi)^2} \left\{ f_0' \left(v_i^0 \partial_j m + v_j^0 \partial_i m + e v_i^0 v_j^0 \Omega  \right) +\hbar f_0'' m v_i^0 v_j^0 + \frac{e \tau}{2(1-i\omega\tau)} f_0' \ve^{\alpha \beta} v_\alpha^0 \partial_\beta\left(v_i^0 v_j^0\right)   \right\} \nonumber \\ &-\frac{e^2 \tau}{1-i\omega \tau} \delta\mu \int\frac{d^2k}{(2\pi)^2} f_0'' v_i^0 v_j^0,
\end{align}
where $\ve^{\alpha \beta}$ is the antisymmetric Levi-Civita tensor (summation over repeated indices is implied) and $\sigma_{ij}^D$ is the zero-field Drude conductivity:
\be  
\sigma_{ij}^D = -\frac{e^2 \tau}{1-i\omega \tau} \int\frac{d^2k}{(2\pi)^2} f_0' v_i^0 v_j^0.
\ee
Here we defined $\partial_i \equiv \partial/\partial k_i$ and suppressed index $\bk$ and argument $\ve^0_\bk$ in $f_0'$, $f_0''$ for brevity. 

The second-order term in $\tau/(1-i\omega\tau)$ originating from the Lorentz force can be rewritten as the full derivative, hence, it vanishes after the integration over the whole Brillouin zone:
\be  
\hbar\int \frac{d^2k}{(2\pi)^2}f_0' \ve^{\alpha \beta} v_\alpha^0 \partial_\beta\left(v_i^0 v_j^0\right) = \int \frac{d^2k}{(2\pi)^2} \ve^{\alpha \beta} \left( \partial_\alpha f_0 \right) \partial_\beta\left(v_i^0 v_j^0\right) = \int \frac{d^2k}{(2\pi)^2} \ve^{\alpha \beta} \partial_\alpha \left[ f_0 \partial_\beta\left(v_i^0 v_j^0\right)  \right] =0, 
\ee 
where we used the identity $\ve^{\alpha \beta} \partial_\alpha \partial_\beta (v_i^0 v_j^0) = 0$ and assumed no singularities in the velocity components.

 Next, the components involving orbital magnetic moment $m_\bk$ can be regrouped using the identity 
\begin{equation}
\hbar f''_0 m v^0_iv^0_j
=
\frac{1}{2}\left[ (\partial_i f'_0) m v^0_j+  (\partial_j f'_0)mv^0_i\right]
=
\frac{1}{2}\left[\partial_i(f'_0mv^0_j)+\partial_j(f'_0mv^0_i)\right]
-\frac{f'_0}{2}\left(v^0_j\partial_im +v^0_i\partial_jm\right)-f'_0 m v^0_{ij},
\end{equation}
where we defined $v^0_{ij} = \partial_i v_j^0 = \partial_j v_i^0$. The first two terms in this expression are full derivatives which vanish upon integration over momenta. Collecting all the non-vanishing contributions in Eq.~\eqref{2Dsigma}, we obtain 
\begin{equation}\label{simplestfullcond}
\sigma^s_{ij}=\sigma^D_{ij}
+\frac{e^2\tau B_z}{\hbar(1-i\omega \tau)}
\int \frac{d^2k}{(2\pi)^2}f'_0
\left\{
\frac{1}{2}(v^0_i\partial_j m_\bk+v^0_j\partial_i m_\bk)
-m_\bk v^0_{ij}+e v^0_iv^0_j\Omega_\bk\right\} + \frac{e^2\tau \delta \mu}{\hbar\left(1-i\omega \tau\right)} \int \frac{d^2k}{(2\pi)^2}f'_0v^0_{ij},
\end{equation}
where we also integrated by parts in the last term.

Equation~\eqref{simplestfullcond} is the most general expression for the symmetric part of the conductivity tensor in two-dimensional systems with the magnetic field along the $z-$axis. If mirror symmetry $m_y$ is additionally present, the term with $\delta \mu$ and $\sigma_{xy}^D$ do not contribute to $\sigma^s_{xy}$, and, in the limit $\omega\to 0$, we reproduce Eq.~\eqref{eq:Sigma} from the main text.

\subsection{Symmetry allowed tight-binding model}

Denote the $d$ orbital creation operators as $d^\dag  = (\hat{d}^\dag_{xz},\hat{d}^\dag_{yz},\hat{d}^\dag_{xy},\hat{d}^\dag_{x^2-y^2},\hat{d}^\dag_{z^2})$ (where we have suppressed the spin component $\hat{d}^\dag_{\ell} = (\hat{d}^\dag_{\ell,\uparrow},\hat{d}^\dag_{\ell,\downarrow})$) which correspond to the following orbital basis
\begin{subequations}
\begin{align}
|d_{xz}\rangle &= \frac{1}{\sqrt{2}}(|2,-1\rangle -|2,1\rangle),\\
|d_{yz}\rangle &= \frac{i}{\sqrt{2}}(|2,-1\rangle + |2,1\rangle),\\
|d_{xy}\rangle &= \frac{i}{\sqrt{2}}(|2,-2\rangle -|2,2\rangle),\\
|d_{x^2-y^2}\rangle &= \frac{1}{\sqrt{2}}(|2,-2\rangle + |2,2\rangle),\\
|d_{z^2}\rangle &= |2,0\rangle,
\end{align}
\end{subequations}
the expressions on the right is written in the $|L,L_z\rangle$ basis.
The onsite trigonal crystal field Hamiltonian has the form
\begin{equation}\label{trigonal_crystal_field_eg}
\hat{H}_{CF}
=d^\dag
\left(
\begin{array}{ccccc}
 \frac{2}{3} & 0 & 0 & \frac{\sqrt{2}}{3} & 0 \\
 0 & \frac{2}{3} & -\frac{\sqrt{2}}{3} & 0 & 0 \\
 0 & -\frac{\sqrt{2}}{3} & \frac{1}{3} & 0 & 0 \\
 \frac{\sqrt{2}}{3} & 0 & 0 & \frac{1}{3} & 0 \\
 0 & 0 & 0 & 0 & 0 \\
\end{array}
\right)d,
\end{equation}
whereas the onsite spin--orbit coupling term has the form
\begin{equation}
\hat{H}_{\text{SO}} = d^\dag \mathbf{\hat{L}}\cdot\mathbf{\hat{S}} d,
\end{equation}
where
\begin{equation}
\hat{L}_x = \left(
\begin{array}{ccccc}
 0 & 0 & i & 0 & 0 \\
 0 & 0 & 0 & -i & -i \sqrt{3} \\
 -i & 0 & 0 & 0 & 0 \\
 0 & i & 0 & 0 & 0 \\
 0 & i \sqrt{3} & 0 & 0 & 0 \\
\end{array}
\right),\qquad
\hat{L}_y
=\left(
\begin{array}{ccccc}
 0 & 0 & 0 & -i & i \sqrt{3} \\
 0 & 0 & -i & 0 & 0 \\
 0 & i & 0 & 0 & 0 \\
 i & 0 & 0 & 0 & 0 \\
 -i \sqrt{3} & 0 & 0 & 0 & 0 \\
\end{array}
\right),\qquad 
\hat{L}_z = \left(
\begin{array}{ccccc}
 0 & -i & 0 & 0 & 0 \\
 i & 0 & 0 & 0 & 0 \\
 0 & 0 & 0 & 2 i & 0 \\
 0 & 0 & -2 i & 0 & 0 \\
 0 & 0 & 0 & 0 & 0 \\
\end{array}
\right),
\end{equation}
and $\mathbf{\hat{S}}$ are written in terms of the standard Pauli matrix operators.
We have set $\hbar=1$.

The triangular lattice is spanned by the lattice vectors $\mathbf{\hat{e}}_1 = \frac{\sqrt{3}}{2}\hat{x} + \frac{1}{2}\hat{y}$ and $\mathbf{\hat{e}}_2 = -\frac{\sqrt{3}}{2}\hat{x} + \frac{1}{2}\hat{y}$, where we have set the lattice constant to unity. We provide the matrix for hopping along the bond $\mathbf{\hat{e}}_1+\mathbf{\hat{e}}_2$ (written in the basis of $d$ orbitals given above): 
\begin{equation}\label{hopping}
[t]_{\mathbf{\hat{e}}_1+\mathbf{\hat{e}}_2} = \left(
\begin{array}{ccccc}
 V_{dd\delta} & 0 & 0 & 0 & 0 \\
 0 & V_{dd\pi} & 0 & 0 & 0 \\
 0 & 0 & V_{dd\pi} & 0 & 0 \\
 0 & 0 & 0 & \frac{1}{4}(V_{dd\delta}+3V_{dd\sigma}) & \frac{\sqrt{3}}{4}(-V_{dd\delta}+V_{dd\sigma}) \\
 0 & 0 & 0 & \frac{\sqrt{3}}{4}(-V_{dd\delta}+V_{dd\sigma}) & \frac{1}{4}(3V_{dd\delta}+V_{dd\sigma}) \\
\end{array}
\right),
\end{equation}
and the other nearest neighbor hoppings can be obtained through translation or threefold rotation (by applying the matrix $\hat{R}_z^{\text{orb}}\left(\frac{2\pi}{3}\right)
\equiv e^{-i\frac{2\pi}{3}\hat{L}_z}$).
Here $V_{dd\sigma}$, $V_{dd\pi}$ and $V_{dd\delta}$ are the standard Slater--Koster parameters \cite{PhysRev.94.1498}. We used the following values in Fig.~2 of the main text:
\begin{equation}
  \lambda_{\text{SO}} =   1.2, \quad V_{dd\sigma}=-1.0, \quad V_{dd\delta}=  0.4, \quad V_{dd\delta}=0, \quad\Delta_{CF}=20.0,
\end{equation}
and a magnetization of norm $|\mathbf{M}|=1.4$.

\section{Landauer--B\"{u}ttiker implementation}

The Landauer--B\"{u}ttiker formalism is employed to calculate $\sigma_{xy}^s$ from the tight-binding model Eq. \eqref{eq:TBHam} in the main text. To that end, we use the Kwant package \cite{Groth2014} to implement our model in the L-shaped device shown in Fig.\ref{fig:LB}(a). The Hamiltonian of the semi-infinite leads is also Eq. \eqref{eq:TBHam}. The parameters are kept the same as those used for calculating $\sigma_{xy}^s$ in Fig. \ref{fig:TightBinding}(d). The width of the device channel and of the leads is taken to be 100 lattice sites, while the length of each arm is 700 sites, resulting in a total device length of $\sim 1400$ sites (see next section for the reasoning behind these values).

Components of the conductivity tensor are not directly obtained in a Landauer--B\"{u}ttiker calculation, but rather they are extracted from inverting the resistivity tensor

\begin{equation}\label{eq:rho_tensor}
\begin{pmatrix} \sigma_{xx} & \sigma_{xy}\\
\sigma_{yx} & \sigma_{yy}
\end{pmatrix} = \frac{1}{\rho_{xx}\rho_{yy}-\rho_{xy}\rho_{yx}}
\begin{pmatrix} \rho_{yy} & -\rho_{xy}\\
-\rho_{yx} & \rho_{xx}
\end{pmatrix}
\end{equation}

\noindent whose components can be computed in the following way. The main outputs from Landauer--B\"{u}ttiker are the transmission probabilities $T_{ij}$ between each pair of leads $i,j$ coupled to a scattering region. Then, imposing the current conditions at each lead (source, drain or floating), the voltages can be obtained with the B\"{u}ttiker formula $I_i = \frac{e}{h} \sum_{j} T_{ji}V_i - T_{ij}V_j$ \cite{Buttiker1986, Datta1997}. With the voltage at each lead and a current $I=I_\text{source}=-I_\text{drain}$, a resistance measurement is given by $R_{ij} = (V_i - V_j)/I$ and hence the resistivity becomes $\rho_{ij} = \frac{w}{d_{ij}} R_{ij}$, with $w$ and $d_{ij}$ being the cross section or width of the device and distance between leads $i$ and $j$, respectively. To minimize the variances produced by the random disorder (see next section), we perform an average over 5 disorder configurations and calculate all components of the resistivity tensor in the L-shaped device. In this way, and taking lead 1 (10) as the source (drain), the resistivities read $\rho_{xx} = \rho_{68}$, $\rho_{yy} = \rho_{24}$, $\rho_{xy} = \rho_{32}$ and $\rho_{yx} = \rho_{89}$.

\subsection{Diffusive transport in Landauer--B\"{u}ttiker calculations}

\begin{figure}[t] 
\centering
\includegraphics[width=0.75\textwidth]{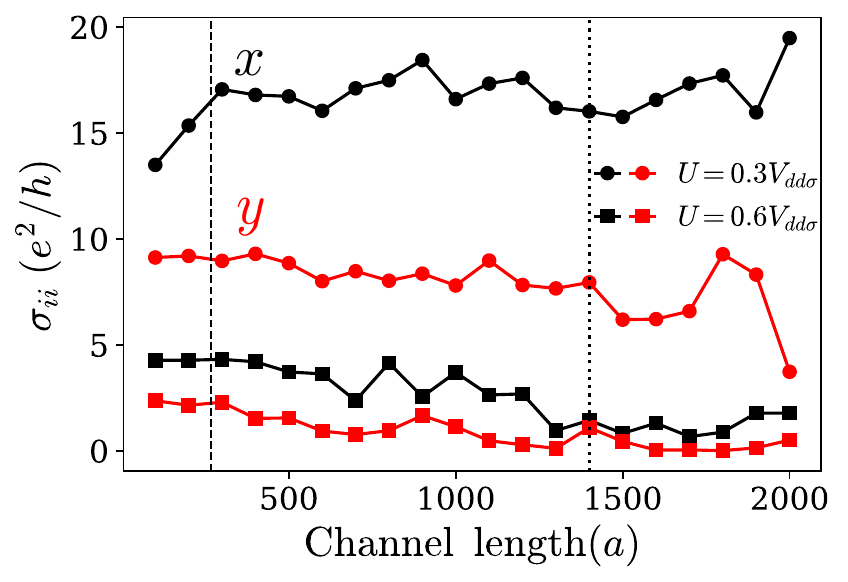}
\caption{Longitudinal conductivity extracted from the two-terminal conductance for two strengths of Anderson disorder and for a device oriented along $x$ (black) and along $y$ (red). 10 disorder averages have been carrid out. The diffusive regime exists for lengths in which the conductivity remains constant. The vertical dashed (dotted) line denotes the chosen inter-lead distance (total length) of the L-shaped device. Importantly, lengths within these two vertical lines fall within diffusive transport.}
\label{fig_Sup_Diffusive}
\end{figure}

Most of mesoscopic transport experiments occur in the diffusive transport regime (as opposed to ballistic or localized transport). In this way, to have diffusive transport in a Landauer--B\"{u}ttiker calculation, we include random Anderson disorder potential $\sum_{\mathbf{r}} U_{\mathbf{r}} d^\dag_{\mathbf{r},\ell,\sigma} d_{\mathbf{r},\ell,\sigma}$ at each site $\mathbf{r}$ uniformly distributed in the interval $[-U/2,U/2]$. Because of the quasi-one-dimensional nature of our finite device, if $U$ is too large, electrons will localize, whereas if $U$ is too small, most of the transport could be ballistic inside our device. Therefore, the choice of $U$ impacts on the choice of the device length and width. To elucidate what is a good value of $U$ so that diffusive transport occurs \textit{throughout} the entire device, we perform the following procedure \cite{Vila2020, VilaThesis}. We make a two-terminal device with the same width as our device of interest (Fig. \ref{fig:LB}(a)) and calculate the two-terminal conductance, $G_{2T}$, for different channel lengths $L$ and fixed $U$. Then, we use the relation $G_{2T}=\sigma \frac{w}{L}$ to obtain the longitudinal conductivity, and plot $\sigma$ as a function of $L$. For an extensive enough range of $L$ values, $\sigma$ will first increase to then remain constant and finally decrease. These three behaviors correspond to ballistic, diffusive and localized transport regimes. Increasing the width will help maximize the range of lengths for which transport is diffusive, and resemble more a mesoscopic two-dimensional system. However, that comes at a higher computational time and memory demand. We found a width of $w=100 a$ ($a$ the lattice constant)  to be big enough for our device, and computationally affordable. Finally, by looking at the scaling of $\sigma$ with $L$, we choose a value of $U$ that allows for a long enough device to probe the physics we are  interested in but still within computational reach. To illustrate this, we plot in Fig. \ref{fig_Sup_Diffusive} the longitudinal conductivity for two different strengths of disorder and for devices oriented along the $x$ and $y$ direction. The Fermi level is taken to be the same as the one used for Fig. \ref{fig:LB}(b) in the main text, i.e. -1.95, and $M_z/M_y = 0.32$ is used. From this plot, it is clear that $\sigma_{xx} \neq \sigma_{yy}$, which is expected due to the broken $c_3$ symmetry. Most importantly, one can see that for the strongest disorder case, the onset of localization appears rather quickly, around $L\sim 500 a$. This is not a good value of $U$, since we cannot build the L-shape device we want within $500a$ if the width of the device and leads is $100a$. In contrast, for $U=0.3V_{dd\sigma}$, the transport is diffusive for both directions at $L \sim 266a$ (see vertical dashed line) and remains constant for long enough lengths. Hence, we choose the inter-lead distance to be at least $L = 266a$. This, together with the width $w=100a$, results in a total length of the L-shaped device of $\sim 1400 a$, which still falls within diffusive transport (vertical dotted line in Fig. \ref{fig_Sup_Diffusive}).

Finally, one also can obtain meaningful quantities such as the mean free path. In the diffusive regime, $G_{2T}=\sigma w/L$, while from quantum transport it is known that $G_{2T}=N l_e/L$ \cite{Datta1997,Foa2014}, where $N$ is the number of propagating modes and $l_e$ the mean free path. Consequently, $\sigma w = N l_e$. Given the leads are described by the same Hamiltonian as the scattering region, $N$ can be extracted from their band structure, whereas $\sigma$ is readily obtained from Fig. \ref{fig_Sup_Diffusive}. This gives us a mean free path of $l_e^x = 31 a$ and $l_e^y = 17 a$ for the $x$ and $y$ directions, respectively.

\section{Density functional theory implementation}

DFT calculations were carried out using the Vienna \textit{ab initio} simulation package (VASP)~\cite{KohnSham1965,Kresse1996,Kresse1999}. The Perdew-Burke-Ernzerhof (PBE) generalized gradient approximation (GGA) was employed for the exchange-correlation potential~\cite{Perdew1997} and the wave functions and pseudopotentials were generated within the projector-augmented wave (PAW) method~\cite{Blochl1994}. 
In our calculations, the monolayer of MnPSe$_3$ was modeled using a periodic slab geometry in a supercell setup. To avoid spurious interactions with periodic images, a vacuum region of 28{\AA} in the out-of-plane direction was used. We employed a plane-wave basis with an energy cutoff of $520$ eV and used a dense k-point grid sampling of $13\times13\times1$ in the BZ according to the Monkhorst-Pack scheme~\cite{Monkhorst1976}. Both the lattice constant and positions of all atoms were relaxed until the force was less than $2$ meV/{\AA}, while keeping the volume fixed. The optimized in-plane lattice constant was found to be $a=b=6.363$ {\AA}, consistent with experimental value ($a=b=6.394$ {\AA})~\cite{Susner2017}. For the electronic self-consistent loop, the total energy convergence criterion was set to $10^{-7}$eV by using GGA (without U). This choice yielded consistent results when compared to calculations with different values of U, as reported in~\cite{Natalya2023}. 

We considered the collinear ferromagnetic (FM) order of the magnetic moment on Mn with spin-orbit coupling (SOC) with its orientation described  as $\mathbf{\hat{m}}=(0, \sin\Theta, \cos\Theta)$, where $\Theta$ varied between $0^\circ$, $+45^\circ$, and $-45^\circ$.

The symmetric and anti-symmetric components of the off-diagonal conductivity tensor were analyzed through explicit calculation of the Fermi velocity and Berry curvature associated with the given Fermi level (chemical potential). We used the Wannier90 code~\cite{Souza2001,Wang2007,Marzari2012} to obtain the Wannier tight-binding Hamiltonian based on maximally localized Wannier functions (MLWFs). To perform the Wannierization procedure, we carried out DFT calculations using a dense k-point grid of $18\times18\times1$. The choice of basis set for the Wannierization included $p_{x,y,z}$ orbitals for the six Se atoms and $d_{xz,yz}$ orbitals for two Mn atoms present in the supercell \cite{Na2023}. This selection allowed us to construct a Wannier tight-binding Hamiltonian of rank $44$. Once we obtained the Wannier tight-binding Hamiltonian, we explicitly calculated the Fermi velocity and the Berry curvature~\cite{Fukui2005,Vanderbilt2006} from this Hamiltonian.

\section{Supporting DFT data for MnPSe$_3$}

In this section, we present DFT calculations to complement the calculation of $\sigma_{xy}^s$ showed in the main text for MnPSe$_3$. 

\begin{figure}[]
\centering
\includegraphics[width=0.75\textwidth]{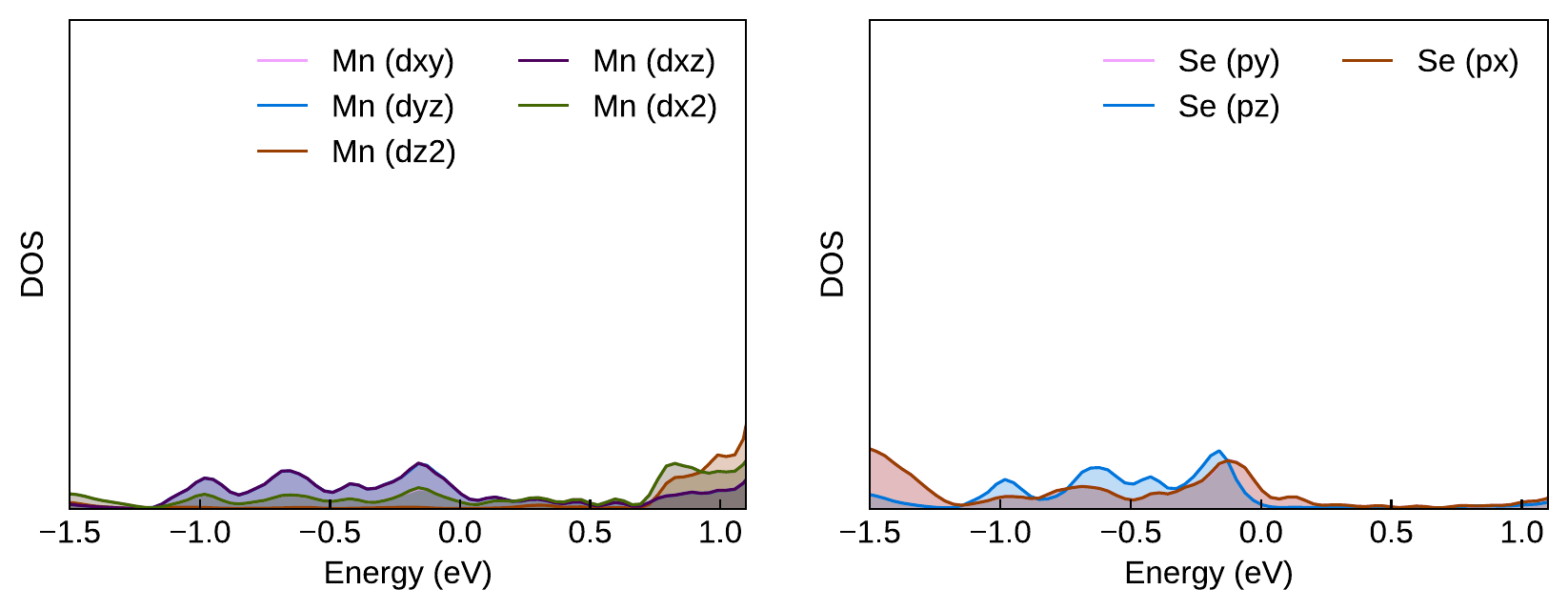}
\caption{Projected density of states of Mn and Se atoms near the Fermi level (set to 0 eV), for MnPSe$_3$ with magnetization orientation along $+ 45\degree$ with respect to the $z$ axis in the $yz$ plane.}
\label{fig_Sup_DOS}
\end{figure}

We argued that in order to observe linear magneto-conductivity, the states at the Fermi level should possess a finite orbital angular momentum even in the presence of a strong crystal field, and demonstrated that with a tight-binding made from $t_{2g}$ orbitals. Therefore, we plot in Fig. \ref{fig_Sup_DOS} the orbital-resolved density of states with magnetization orientation at $+ 45\degree$ with respect to the $z$ axis in the $yz$ plane. As expected from the nonzero $\sigma_{xy}^s$ numerically computed in Fig. \ref{fig:DFT} in the main text, MnPSe$_3$ shows considerable weight of $t_{2g}$ orbitals ($d_{xy}$, $d_{xz}$, $d_{yz}$) near the Fermi level, thus validating our understanding of the importance of having a finite orbital angular momentum.    

\begin{figure}[] 
\centering
\includegraphics[width=0.75\textwidth]{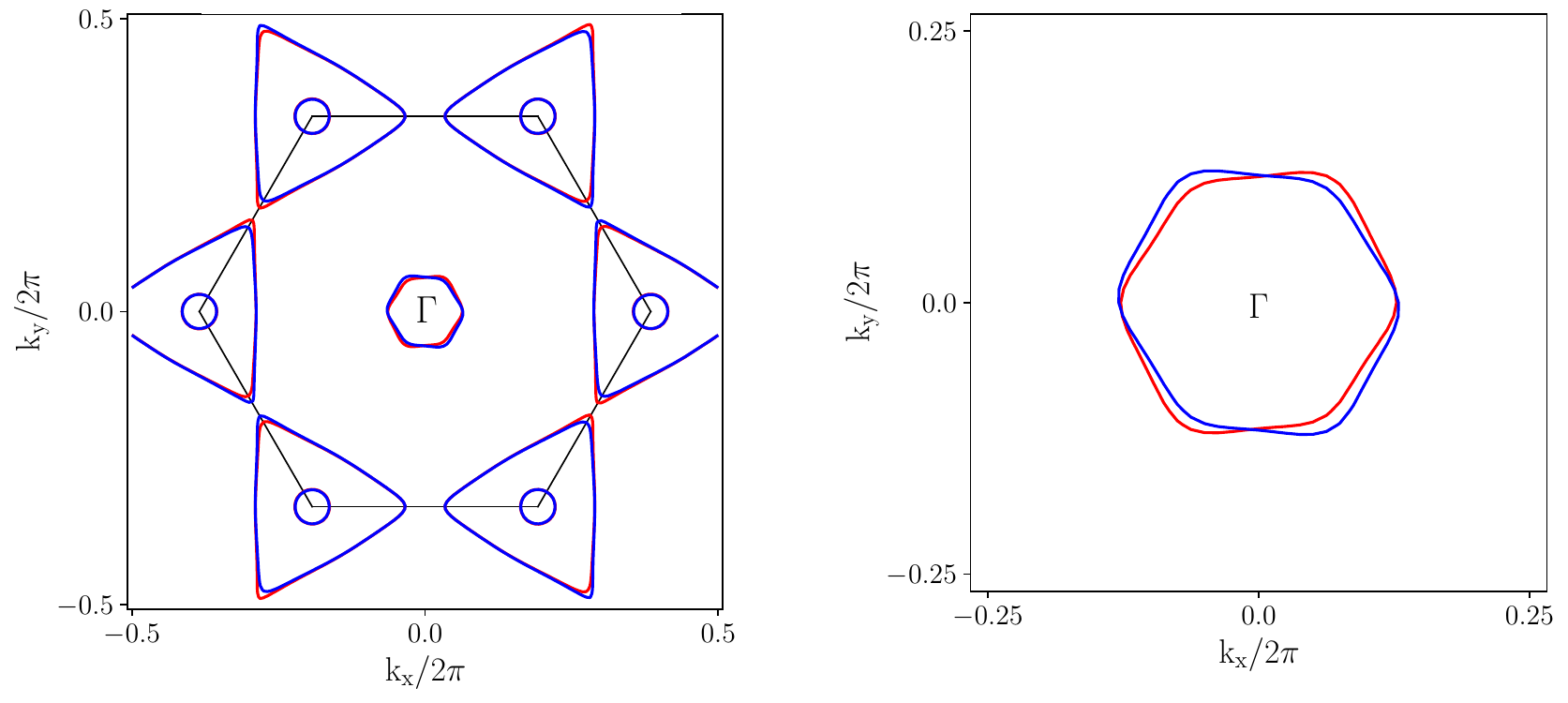}
\caption{Fermi surface at energy $-110$ meV with respect to the valence band maximum for MnPSe$_3$ with magnetization orientation along $+ 45\degree$ (red) and $- 45\degree$ (blue). Left (right) panel show the Fermi surface in the whole Brillouin zone (around the $\Gamma$ point).}
\label{fig_Sup_FS}
\end{figure}

Another property that we argued is tied to the appearance of $\sigma_{xy}^s$ is the rotation of the Fermi surface with respect to the principal axis. Fig. \ref{fig_Sup_FS} displays the Fermi surface at an energy -110 meV with respect to the valence band maximum for magnetizations $\pm 45\degree$. The Fermi surfaces have pockets around the Brillouin zone corners (K points) and round the $\Gamma$ point. In the K points, the Fermi surface rotation between the two magnetization configurations is not very clear, but each Fermi surfaces clearly shows the breaking of both $c_3$ and $m_y$ symmetries. On the other hand, the Fermi surfaces at the $\Gamma$ point reveal opposite rotation in the $xy$ plane. This result, together with Fig. \ref{fig_Sup_DOS} and Fig. \ref{fig:DFT} in the main text, demonstrate the correlation between orbital angular momentum, Fermi surface rotation and nonzero $\sigma_{xy}^s$ in realistic materials.

\end{document}